\documentclass[prb,preprint,aps,superscriptaddress,longbibliography]{revtex4-1}
\usepackage[]{color}
\usepackage{appendix}
\usepackage{graphicx, epstopdf}
\usepackage{subcaption}
\usepackage{float}
\usepackage[utf8]{inputenc}
\usepackage{textgreek}
\usepackage{amsmath}
\usepackage{amsfonts}
\usepackage{amssymb}
\usepackage{bm}
\usepackage{caption}
\usepackage{hyperref}
\usepackage{xr}
\usepackage{lineno}
\usepackage{natbib}
\usepackage{soul}
\usepackage{xr}
\usepackage{lineno}
\usepackage{epsfig}
\usepackage{tabularx}
\begin{document}

\title{Phonon-informed Crystal Structure Classification via Precision-Adaptive ResNet-based Confidence Ensemble}

\author{Hongyu Chen}
\thanks{These authors contributed equally to this work.}
\affiliation{College of Future Information Technology, Fudan University, Shanghai 200433, China}
\affiliation{Key Laboratory of Micro and Nano Photonic Structures (MOE), Fudan University, Shanghai 200433, China}

\author{Mengyu Dai}
\thanks{These authors contributed equally to this work.}
\affiliation{College of Future Information Technology, Fudan University, Shanghai 200433, China}
\affiliation{Key Laboratory of Micro and Nano Photonic Structures (MOE), Fudan University, Shanghai 200433, China}

\author{Hongjiang Chen}
\thanks{These authors contributed equally to this work.}
\affiliation{College of Future Information Technology, Fudan University, Shanghai 200433, China}
\affiliation{Key Laboratory of Micro and Nano Photonic Structures (MOE), Fudan University, Shanghai 200433, China}

\author{Ruilin Liu}
\affiliation{College of Future Information Technology, Fudan University, Shanghai 200433, China}
\affiliation{Key Laboratory of Micro and Nano Photonic Structures (MOE), Fudan University, Shanghai 200433, China}

\author{Xiaole Tian}
\affiliation{College of Future Information Technology, Fudan University, Shanghai 200433, China}
\affiliation{Key Laboratory of Micro and Nano Photonic Structures (MOE), Fudan University, Shanghai 200433, China}

\author{Ruixiao Lian}
\affiliation{College of Future Information Technology, Fudan University, Shanghai 200433, China}
\affiliation{Key Laboratory of Micro and Nano Photonic Structures (MOE), Fudan University, Shanghai 200433, China}

\author{Yuqian Zhang}
\affiliation{College of Future Information Technology, Fudan University, Shanghai 200433, China}

\author{Xia Cai}
\affiliation{College of Future Information Technology, Fudan University, Shanghai 200433, China}

\author{Wenwu Li}
\affiliation{College of Future Information Technology, Fudan University, Shanghai 200433, China}

\author{Hao Zhang}
\email{zhangh@fudan.edu.cn}
\affiliation{College of Future Information Technology, Fudan University, Shanghai 200433, China}
\affiliation{Key Laboratory of Micro and Nano Photonic Structures (MOE), Fudan University, Shanghai 200433, China}
\affiliation{State Key Laboratory of Photovoltaic Science and Technology, Fudan University, Shanghai 200433, China}

\begin{abstract}
Accurate description of crystal structures is a prerequisite for predicting the physicochemical properties of materials. However, conventional X-ray diffraction (XRD) characterization often encounters intrinsic bottlenecks when applied to complex multiphase systems, necessitating the integration of complementary optical measurement. In this study, we developed a multi-descriptor framework by integrating key parameters including space groups, Pearson symbols, and Wyckoff sequences, to categorize the dataset of over 19,000 crystals into several dozen structural prototypes. Then, an accuracy-adaptive ensemble network based on residual architectures was implemented to capture structural ``fingerprints" within phonon vibration modes and Raman spectra. The ensemble algorithm demonstrates exceptional robustness when processing various crystals of varying lengths and quality. This data-driven classification strategy not only overcomes the reliance of traditional characterization on ideal data but also provides a high-throughput tool for the automated analysis of material structures in large-scale experimental workflows.

\vspace{2mm}

\noindent{\textcolor{black}{\textbf{KEYWORDS:} Phonon dispersions; Crystal structures; Resnet-based confidence ensemble; Structure-properties relationship}}

\end{abstract}

\flushbottom
\maketitle

\thispagestyle{empty}

\section{Introduction}

The vast diversity of materials underpins modern civilization, enabling technologies ranging from energy storage and carbon capture to semiconductor-based microprocessing\cite{ling2022review,hu2023machine,choubisa2023interpretable}. As crystalline solids constitute the structural foundation of most materials, estimating their structures or identifying their structure information in 3D space is central to understanding and predicting their physical and chemical properties, which greatly influence their applications to various sciences, such as the design of drugs, batteries, micro-chips, and catalyst\cite{butler2018machine}.
A crystal structure can be represented as the infinite periodic arrangement of atoms in the Euclidean space, and the smallest repeating unit called a unit cell can be defined by a triplet $\mathcal{M}=[\mathbf{A},\mathbf{X},\mathbf{L}]$, where $\mathbf{A}=[a_1,a_2,...,a_N]$ are the list of chemical species, $\mathbf{X}=[\mathbf{x_1},\mathbf{x_2},...,\mathbf{x_N}]\in \mathbb{R}^{3\times N}$ are the Cartesian coordinates of the atoms, and $\mathbf{L}=[\mathbf{l_1},\mathbf{l_2},\mathbf{l_3}]\in \mathbb{R}^{3\times3}$ denotes the lattice matrix containing three basic vectors describing the periodicity of the crystal. $N$ is the number of atoms in the unit cell. 

Essentially, the periodic arrangement of crystal structures is constrained by three fundamental symmetries, i.e. permutation invariance of atomic indices, $O(3)$ rotational symmetry, and translational symmetry, with the latter two jointly defining the space group symmetry $G$\cite{dresselhaus2007group}. In 3D space, there exist 230 space-group symmetries\cite{hahn1983international}, which restrict the admissible values of the lattice matrix $\mathbf{L}$ and the positions of atoms. The symmetry operations $g\in G$ that leave a given atom $\mathbf{X_i}$ invariant, that is, mapping symmetry-equivalent configurations of the crystal onto themselves, define the site-symmetry subgroup $G_i\subseteq G$, given by, $G_i=\{g\in G|g\cdot\mathbf{X_i}=\mathbf{X_i}\}$. Atoms located at $\mathbf{X_i}$ generate equivalent positions $\{g_s\mathbf{X_i}+\mathbf{\tau_s}\}_{s=1}^{n_s}$, where $g_s\in G_i$, $\mathbf{\tau_s}$ is the translational vector, and $n_s$ is the multiplicity of the symmetry-equivalent positions. The multiplicity $n_s$ counts the number of symmetry-equivalent positions generated by the site-symmetry operations, and all of them should be occupied by identical type of atoms to uphold the space group symmetry. These site symmetry points can be grouped into Wyckoff positions (WPs)\cite{wyckoff1922analytical}, which encompass all points whose site-symmetry groups are conjugate subgroups of the full space group\cite{kantorovich2004quantum}. WPs of a given space group are labeled by alphabetical letters, typically ordered by decreasing site symmetry, with ``a” denoting the position of highest-order site symmetry in the space group. 
With these considerations, a crystal data can be further deliberately defined as $\mathcal{C}=[\mathbf{W},\mathbf{A},\mathbf{X},\mathbf{L^\prime}]$\cite{cao2025space}, with $\mathbf{W}=[w_1,w_2,...,w_n]$ the Wyckoff letters, $\mathbf{L^\prime}=[a,b,c,\alpha,\beta,\gamma]$ the lattice parameters of unit cell, and $n$ the number of symmetrically inequivalent atoms in the unit cell.

With the advent of automated workflows and high-throughput experimentation in materials science\cite{stein2019progress}, the capacity to synthesize and screen a large number of materials has increased dramatically, 
and rapid and reliable structural characterization is indispensable. Traditional experimental techniques for structure determination include scanning electron microscopy (SEM)\cite{zhou2006fundamentals}, transmission electron microscopy (TEM)\cite{williams1996transmission}, X-ray diffraction (XRD)\cite{hauptman1986direct,patterson1935direct,karle1964application,green1954structure}, and etc. SEM and TEM produce high-resolution images by interacting electron beams with samples, revealing their morphology and crystallographic orientation at the microscale. XRD, based on Bragg diffraction, measures peak positions and intensities to provide average information on crystal lattices and phase composition. These techniques serve as the state-of-the-art tools for microstructure, phase, and crystal structure analysis in materials discovery.
%
Despite its widespread usages, XRD, particularly powder XRD, faces significant limitations when applied to complex or multiphase systems\cite{harris1994crystal}. Overlapping reflections, weak diffraction signals, and structural similarities among polymorphs often hinder accurate phase and structure description\cite{rietveld1969profile}. These challenges are further exacerbated in high-throughput synthesis, where a large number of polycrystalline or microcrystalline samples are produced.

As a complementary technique, Raman spectroscopy overcomes several of these limitations by probing local regions, thin films, and small particles that are difficult to characterize using XRD\cite{smith2019modern}. Moreover, Raman spectra directly reflect phonon vibrations and lattice dynamics, which encode information about atomic arrangements and crystal symmetry. Consequently, Raman spectroscopy provides additional structural fingerprints that supplement XRD measurements, thereby improving the robustness and reliability of structural description. Recent studies combining Raman spectroscopy with machine learning or deep learning have demonstrated notable success in automatic material or mineral classification\cite{lussier2020deep,jinadasa2021deep}, including the discrimination of polymorphs or mineral species in complex or mixed systems. However, many existing works focus primarily on identifying chemical composition or mineral identity rather than explicitly targeting crystal structure type classification.


In this work, we introduce the Precision-Adaptive ResNet-based Confidence Ensemble (PARCE), a model built on convolutional residual neural networks (ResNet)\cite{he2016deep}, to automate the identification of the most probable structure type of a material from its $\Gamma$-point phonon sprctra. The applicability and potential of PARCE are demonstrated on a dataset of 40339 materials extracted from the \textit{Materials Project} (MP) database\cite{jain2013commentary}. The overall workflow follows a ``multi-level clustering + deep learning'' strategy to map phonon spectra to structure-type labels, 
which comprises two stages: a data-processing stage based on multi-level clustering and a model-training-and-ensembling stage based on deep learning. In the data-processing stage, we first apply the K-means clustering algorithm to structural parameters extracted from $\mathcal{C}$, that is, space group, Pearson symbol, Wyckoff sequence, $c/a$ ratio, and $\beta$ angle, to group materials according to similar structural attributes, which serves as an approximate representation of a structure type. Next, for each cluster, $\Gamma$-point phonon spectra are standardized and truncated, and a cluster-average spectrum is computed to obtain representative spectra. Finally, using the representative spectra as features, we conduct an additional round of clustering with the affinity propagation algorithm to generate the category partitions required for supervised learning. Each affinity-propagation--derived category, together with its constituent materials, forms a corresponding training subset, thereby assembling the training dataset.
In the model-training-and-ensembling stage, we adopt the ResNet Confidence Networks (RCNet) architecture to learn the mapping between phonon spectral features and structure types\cite{chen2024crystal}, with the network outputting candidate structure types along with their confidence scores. Based on validation performance, the top-performing RCNet models are ensembled within PARCE to produce the final prediction of the material's structure classification from the input phonon spectrum.

\section{Results and discussion}

\subsection{Clustering Strategy for Inorganic Crystal Materials}

\begin{figure}[ht!]
\centering
\includegraphics[width=0.9\linewidth]{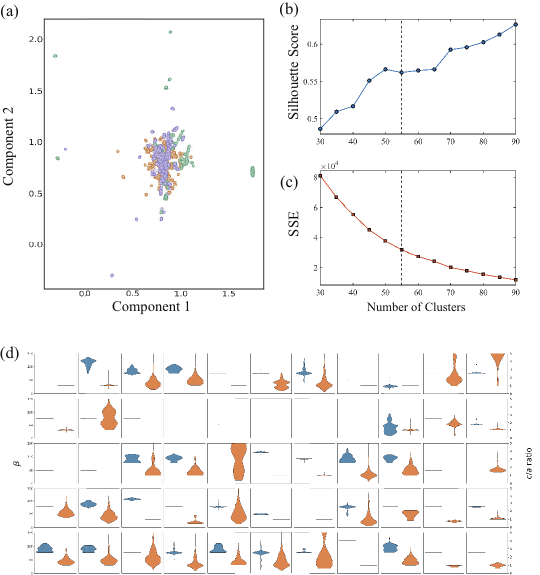}
\caption{(a) 2D UMAP-projected K-means clustering results of 55 clusters categorized based on the (12345) parameter combination. (b) Silhouette score as a function of the number of clusters. (c) Sum of squared errors (SSE) as a function of the number of clusters. (d) Statistical distributions of the crystallographic parameters of $\beta$ angle ($^\circ$) (blue violin) and $c/a$ ratio (brown violin) across 55 clusters.}
\label{fig:cluster}
\end{figure}

To build the hierarchical classification for the inorganic MP materials with physical rationality\cite{jain2013commentary},
~the clustering strategy we proposed can be primarily divided into two classification stages. The first stage, referred to as the clustering of inorganic materials, aims to classify over ten thousand MP crystal materials according to their fundamental crystal structure features. Specifically, five key crystallographic parameters including space group, Pearson symbol, Wyckoff sequence, $c/a$ ratio, and $\beta$ angle, are calculated and extracted, as these descriptors are widely used to characterize and classify inorganic crystals in the ICSD database\cite{allmann2007introduction}. Subsequently, the K-means clustering algorithm is applied to categorize the inorganic MP materials based on different combinations of these parameters\cite{scikit-learn}, generating $K$ ($K=16$ here) distinct groupings, each comprising three or more of the five key parameters. Essentially,
%
the resulting clusters for a specific parameter group $k$ ($k=1,\dots,K$) contain materials that are similar in terms of their structural characteristics. 
To assess the quality and reliability of the K-means clustering results, both the Silhouette Score and the sum of squared errors (SSE) are employed\cite{scikit-learn,nainggolan2019improved}. The Silhouette Score quantifies cluster cohesion and separation, with higher values indicating that the materials are more appropriately grouped according to their combined crystallographic parameters. By using the Silhouette Score, the optimal clustering scheme can be identified as shown in Figure~\ref{fig:cluster}(b), ensuring both physical interpretability and statistical robustness. The SSE, defined as the sum of squared Euclidean distances between each point and its nearest centroid, serves as a joint criterion for determining the optimal number of clusters, as indicated by the elbow point of the SSE curve, as shown in Figure~\ref{fig:cluster}(c).

\begin{figure}[ht!]
\centering
\includegraphics[width=0.9\linewidth]{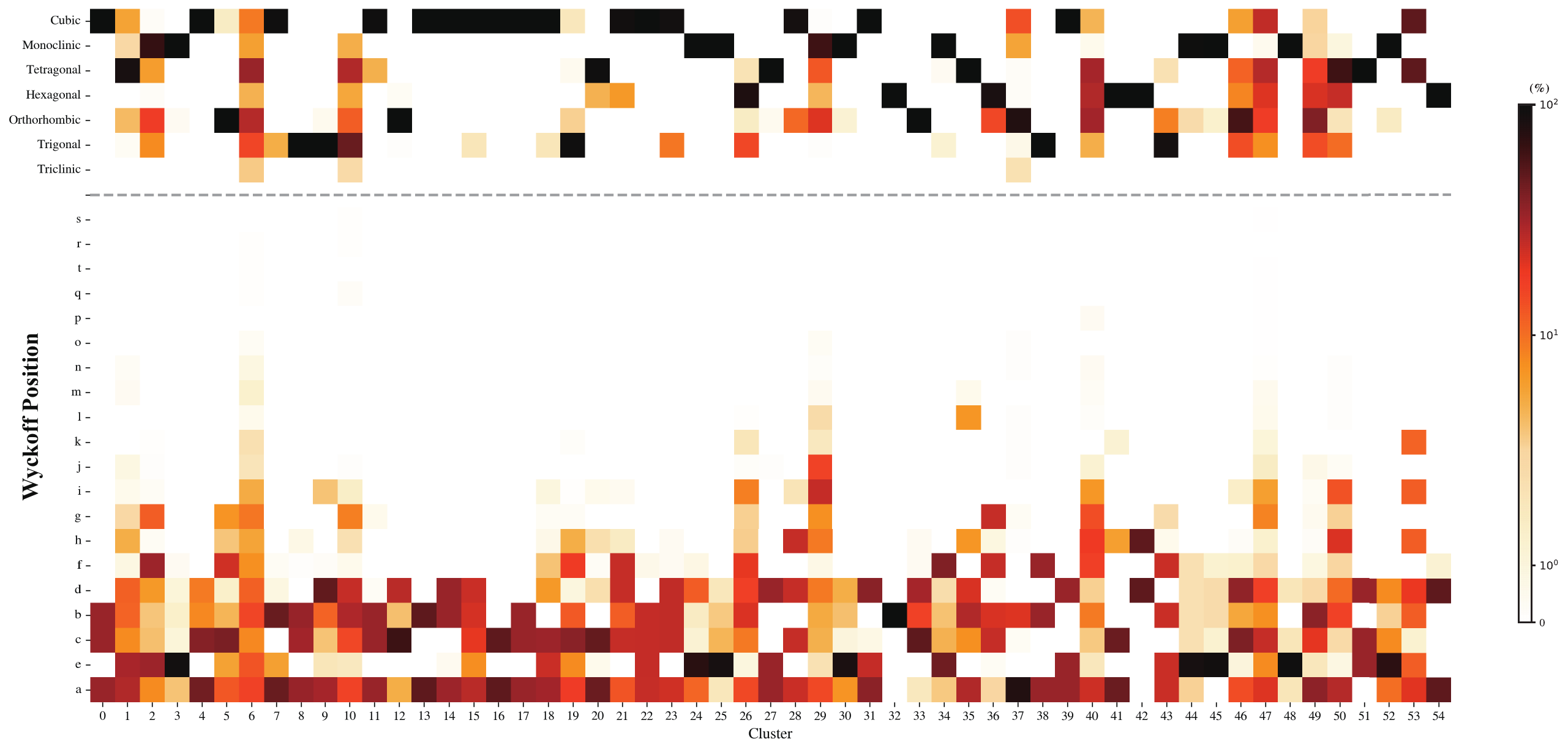}
\caption{Statistical distributions of the crystallographic parameters of  crystal system (above) and Wcykoff position (below) of structures across 55 clusters.}
\label{fig:wcykoff}
\end{figure}

Based on the joint criteria of the Silhouette Score and SSE, $L$ were selected from $K$ possible candidates (in this case, $L=3$). The three optimal combinations, corresponding to $l=1,2,3$, were identified as (124), (125) and (12345). In these combinations, the digits 1–5 refer to the space group, Pearson symbol, Wyckoff sequence, $c/a$ ration and angle $\beta$, respectively.
The Silhouette Scores for these combinations are  0.774, 0.836, and 0.656, and the corresponding SSE values are roughly equal to 3.58$\times10^5$, 2.61$\times10^5$, and 3.17$\times10^5$, leading to optimal cluster numbers $W_l$ of 50, 55, and 55, respectively.
For clear physical interpretability and without loss of generality, the clustering results for over 19,000 MP materials based on the (12345) combination were considered and shown in Figure~\ref{fig:cluster}(a),  in which the high-dimensional data have been projected into a two-dimensional space using the Uniform Manifold Approximation and Projection (UMAP) method for clarity \cite{mcinnes2018umap}. The UMAP results indicate that the clusters are sufficiently well separated, demonstrating the rationality of the clustering. 
The distributions of the $\beta$ angles and $c/a$ ratios for structures in respective clusters are shown in Figure~\ref{fig:cluster}(d), and the corresponding distributions of Wyckoff positions and space group numbers grouped by crystal system, are shown in Figure~\ref{fig:wcykoff}.

Due to the structural similarity of the crystal materials within the same structure cluster, their corresponding physical properties are also expected to be similar. Thus, in the second stage, the phonon frequencies at the $\Gamma$ point for all materials within a specific cluster are averaged to serve as the representative phonon frequencies for that structure cluster. For a given clustering scheme with $W_l$ ($l=1,...,L$) structure clusters, as mentioned above, this process yields $W_l$ averaged phonon frequencies at the $\Gamma$ point. In addition, since the number of atoms in primitive cells differs, the lengths of phonon frequencies varies across materials. To standardize them before averaging, a truncation and zero-padding scheme is applied.
~The upper bound was determined statistically to cover 90\% of MP materials, which corresponds to 33 frequencies for inorganic MP materials, and the lower bound was set to 3, accounting for the lowest optical branch. For simplicity and without loss of generality, a range of length choices from the lower bound to the upper bound, with an interval of 6, was considered, ultimately yielding $M$ ($M = 11$ here) distinct datasets of standardized phonon frequencies at the $\Gamma$ point. Consequently, for a given clustering scheme with generated $W_l$ structure clusters, there will be $M$ corresponding sets of averaged phonon frequencies at the $\Gamma$ point.

Furthermore, for the $W_l$-averaged phonon-frequency data at the $\Gamma$ point, truncated by a cut-off $P_{m}$ ($m = 1, \dots, M$) relative to the phonon-frequency length, we employed the affinity propagation method\cite{scikit-learn} to categorize the data, along with the corresponding materials, into $N_{lm}$ ($l=1,\dots,L; m = 1, \dots, M$) categories. As listed in Table~\ref{tab:cluster_form}, for the three optimal parameter combinations with the highest Silhouette Scores and appropriate elbow points on the SSE curves, i.e., (124), (125), and (12345), the affinity propagation results indicate that the number of categories is ranging from 3 to 5. 
This categorization provides the foundation for constructing a deep-learning model that maps the input phonon frequencies at the $\Gamma$ point to the corresponding structure clustering information, as illustrated in Figure~\ref{fig:parce_workflow}. Consequently, the total number of constituent deep-learning models, $N_\Phi$, equals to the total number of categories listed in Table~\ref{tab:cluster_form}, namely,

\begin{equation}
N_\Phi=\sum_{l=1}^L\sum_{m=1}^M{N_{lm}},
\end{equation}

Where $l$ represents the index of parameter combinations , $M$ denotes the truncation index, and $N_{lM}$ represents the corresponding category associated with the $l$-th parameter combination and the $M$-th truncation.

\subsection{PARCE Model: Architecture}

\begin{figure}[ht!]
    \centering
    \includegraphics[width=1\textwidth]{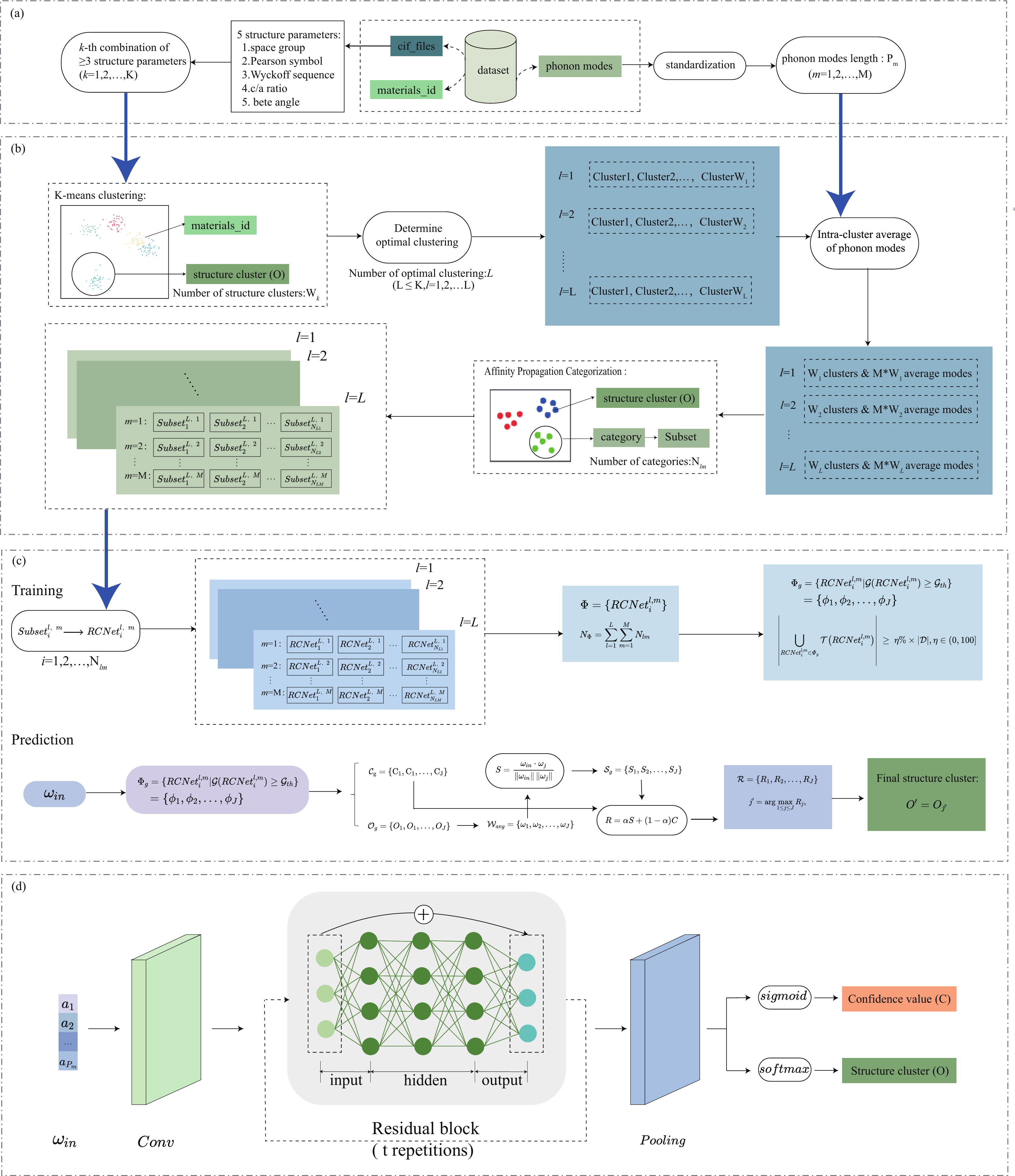}
    \captionsetup{format=plain, justification=raggedright, singlelinecheck=false}
    \caption{Overall Workflow of the PARCE model. (a) Initial preparation of material structures and phonon modes; (b) Clustering-based processing of the initial data; (c) Workflow of RCNet Ensemble, Training, Filtering, and Final Prediction; (d) Single RCNet architecture}
    \label{fig:parce_workflow}
\end{figure}

As mentioned above, the structure clusters for crystal materials are predefined by applying unsupervised clustering algorithms such as K-means, UMAP and so on, to classify crystals materials in the MP dataset, based on a composite key crystallographic parameters including space group, Wyckoff position sequences, Pearson symbols, $c/a$ ratios, and $\beta$ angles. These structure clusters, which contain crystal structure information in a statistical sense, can be used as task labels for classification tasks to classify new crystal structures, and they can also be used to predict the structural information of unknown crystals based on their phonon frequencies data, by mapping the statistical patterns of phonon frequencies to the corresponding structure cluster. 
This task is accomplished by the proposed Precision-Adaptive ResNet-based Confidence Ensemble (PARCE) model, with the architecture shown in Figure~\ref{fig:parce_workflow}. The design of the model is based on the physical hypothesis that the phonon spectrum of a material encodes sufficient information of its crystal structure by a surjective relationship. Unlike traditional classification methods that rely on explicit geometric or symmetry data, our approach requires only phonon frequency input, making it highly adaptable to cases where structural information is missing or incomplete. The model outputs the predicted structure cluster for phonon-frequency input, providing a practical tool for retrieving structural information in data-limited or high-throughput screening scenarios.

Overally, as shown in Figure~\ref{fig:parce_workflow}, the PARCE model consists of four parts: (a) dataset preparation (Figure~\ref{fig:parce_workflow}(a)); (b) processing the dataset to generate batch subsets suitable for model training (Figure~\ref{fig:parce_workflow}(b)); (c) the training and prediction process of the deep-learning integration module (Figure~\ref{fig:parce_workflow}(c)); and (d) the architecture of a single RCNet\cite{he2016deep,devries2018learning,chen2024crystal}, described in detail in~\ref{sec:rcnet} and illustrated in Figure~\ref{fig:parce_workflow}(d).

As mentioned above, based on the high-throughput \textit{ab-initio} calculations of phonon dispersions of materials in the MP dataset, with the methods described in~\ref{sec:ab}, 
the structural information, phonon frequencies at the $\Gamma$-point or Raman frquencies of materials were collected as the database. By using the five key crystallographic parameters, including space group, Wyckoff position sequences, Pearson symbols, $c/a$ ratios, and $\beta$ angles (or some of them), all the collected materials can be classified into $W_k$ ($k=1,..,K$) clusters using the K-means clustering algorithm. Obiviously, the resulting number of clusters $W_k$ depends on the choice of the combination of crystallographic parameters.
Due to the necessity of using standardized data for model training, the phonon frequencies are normalized to ensure consistent lengths. To retain certain features related to the phonon frequency length, a truncation is applied to the normalized phonon frequencies. Subsequently, each truncated segment is used for training. The truncation length is denoted as $P_m (m = 1, 2, \dots,M)$. As shown in Figure ~\ref{fig:parce_workflow}(a), the preliminary dataset, including both material structural information and phonon data, is prepared.

The ensemble model requires training on different combinations of structural parameters, varying phonon frequency truncation lengths, and combinations with distinct phonon frequency characteristics. The method for forming these specific data subsets is illustrated in Figure ~\ref{fig:parce_workflow}(b). Considering the clustering evaluation metrics, $L$ parameter combinations ($ L \leq K $) are selected from the $K$ possible structural parameter combinations, with the chosen $L$ combinations demonstrating relatively superior clustering performance. After clustering with the $l$-th ($ l = 1, 2, \dots, L $) parameter combination, $ W_l $ clusters are formed. To extract the phonon feature information of the structural clusters, the phonon frequencies of all materials within each cluster are averaged, based on the previously applied phonon frequency truncation. The \( l \)-th parameter combination results in \( W_l \) clusters, yielding a total of \( M \times W_l \) averaged phonon frequencies.

For each parameter combination, the clusters corresponding to each truncation are subjected to affinity propagation clustering. The purpose of affinity propagation clustering is to group structure clusters with similar phonon frequency characteristics into the same category. For the \( l \)-th parameter combination and phonon frequency truncation \( P_m \), the corresponding cluster set can be divided into \( N_{lm} \) categories. Each category represents a data subset that includes various materials, where each material is characterized by its phonon frequency data at truncation \( P_m \) and the corresponding structure cluster. For \( L \) parameter combinations and \( M \) truncations, the affinity propagation clustering procedure generates a total of \(N_{subset} \) data subsets, where \(N_{subset}=N_\Phi=\sum_{l=1}^L\sum_{m=1}^M{N_{lm}}\)

As shown in the training section of Figure ~\ref{fig:parce_workflow}(c), the ensemble model undergoes preliminary training, selection, and evaluation processes. The data subset \(Subset^{l,m}_i\) corresponds to the training of \(RCNet^{l,m}_i\). Herein, for a certain structure parameter conbination \(l\),a deep-learning integration module with $M$ blocks and each block contains $N_{lm}$ ($m=1,\dots M$) branches of ResNet Confidence Network (RCNet), which can predict structure clustering information and its confidence simultaneously, is proposed to build a classification model that maps the phonon frequencies at the $\Gamma$ point of materials to their corresponding structure clustering information. The single RCNet model, with the architecture shown in Figure~\ref{fig:parce_workflow}(d) and the model algorithm described in ~\ref{sec:rcnet}, outputs the structure cluster (O) and the confidence value (C).

Due to the complexity of the intrinsic physical correlation between crystal structures and phonon frequencies for different materials, as well as the number of materials in different clusters, the precision of each RCNet differs significantly, as shown in Figure~\ref{fig:performance}. Therefore, a set of high-precision RCNets can be selected to form a complete module for accurately mapping phonon frequencies to structure clusters. As shown in Figure~\ref{fig:parce_workflow}(c), a set of RCNets whose validation accuracy ($\mathcal{G}$) exceeds a predefined threshold ($\mathcal{G}_{th}$) was selected, and the resulting subset is illustrated in Figure~\ref{fig:parce_workflow}(c) as $\Phi_g$. These selected RCNet blocks can be formally expressed as
$\Phi_g = \left\{ RCNet^{l,m}_i \mid \mathcal{G}(RCNet^{l,m}_i)) \geq \mathcal{G}_{\mathrm{th}} \right\}$, where the index $i \in \{{1,2,\dots,N_{lm}}\}$ specifies the individual RCNet 
within the $m$-th truncation group and $l$-{th} key-parameter clusters.Further analysis reveals that the materials included in this selected subset account for $\eta\%$ ($\eta \in (0,100]$, e.g., 90\%) of the entire dataset, which to some extent demonstrates the rationality of the selection strategy.  For model training, hyperparameter optimization is required in the selection of the RCNet subset. 

The procedure of employing the integrated ensemble $\Phi_g$ to predict the structure cluster corresponding to a given phonon spectrum is illustrated in the prediction section of Figure~\ref{fig:parce_workflow}(c). Specifically, the phonon spectrum is normalized to the appropriate length $P_m$ required by each $RCNet^{l,m}_i$, after which all networks $\phi_j(j=1,2,\dots,J)$ belonging to $\Phi_g$ are utilized to predict the structure cluster $O_j$ together with the associated confidence value $C_j$, thereby yielding the sets $\mathcal{O}_g=\{O_1,O_2,\dots,O_J\}$ and the sets $\mathcal{C}_g=\{C_1,C_2,\dots,C_J\}$. For each predicted structure cluster $O_j$, the corresponding average phonon spectrum can be obtained, and its cosine similarity with the normalized input spectrum is subsequently computed to form the set $\mathcal{S}_g=\{S_1,S_2,\dots,S_J\}$. The $\mathcal{S}_g$ is then linearly combined with the confidence values to yield the reliability score $R_j$, defined as $R_j=\alpha S_j+(1-\alpha)C_j$, where $\alpha$ denotes the coupling parameter. The final prediction result is determined by the maximum $R_j$, which corresponds to the structure cluster $O_j$ associated with the selected $\phi_j$.

\subsection{PARCE Model: Model Performance and Examples}

The two-stage clustering pipeline in the PARCE workflow as shown in Figure~\ref{fig:parce_workflow}(a,b) yielded 124 distinct datasets for the combinations of (124), (125) and (12345) with high Silhouette Scores described above, as listed in Table~\ref{tab:cluster_form}. Each dataset is used to train a separate RCNet model for 500 epochs, generating respective training and validation curves for both loss and accuracy over epoch times, and the models with excellent accuracies are selected to form an ensemble, which is then used to construct the integrated model for the complete dataset, as shown in Figure~\ref{fig:parce_workflow}(b,c). 

\begin{figure}[ht!]
    \centering
    \includegraphics[width=0.9\linewidth]{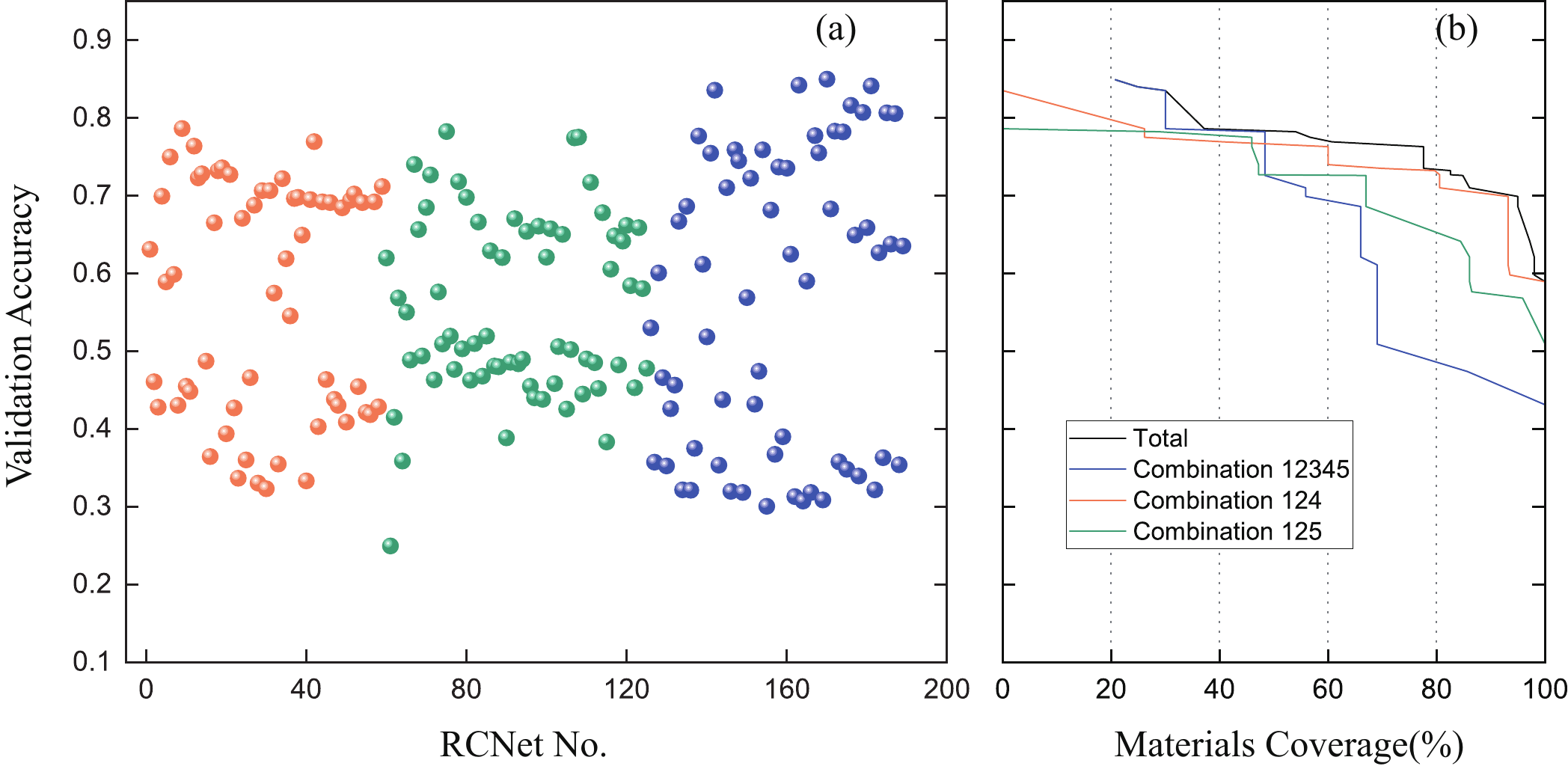}
    \caption{(a) Validation accuracy distribution across RCNet configurations based on the combinations of (124) (orange), (125) (green), and (12345) (blue). (b) Materials coverages for respective and total combinations of key pxarameters.}
    \label{fig:performance}
\end{figure}

To provide a comprehensive overview of the results, we statistically aggregated the performance of all 189 constituent RCNet models, and their validation accuracies and cumulative coverages over the whole dataset are show in Figure~\ref{fig:performance}(a,b), respectively.
%
%
~The curves in Figure~\ref{fig:performance}(b) represent the cumulative percentage of the training dataset with respect to the RCNet models included for specific combinations, such as the (124) combination shown by the orange line. As the RCNet models for the (124) combination are sorted by their accuracies, the proportion of cumulative training data corresponding to these models increases. This continues until the accuracy reaches approximately 62.4\% (covering 14 RCNet models for the (124) combination), at which point the training data encompasses the 90\% of entire dataset. The accuracy thresholds for the (125) combination, denoted by the green line, and the (12345) combination, denoted by the blue line, are 61.8\% and 61.5\%, respectively, and the numbers of corresponding involved RCNet models are 45 and 38, respectively. The above three combinations of structural parameters and their corresponding thresholds were selected and evaluated on the test set to obtain the highest accuracy.



\begin{figure}[ht!]
    \centering
    \includegraphics[width=0.9\linewidth]{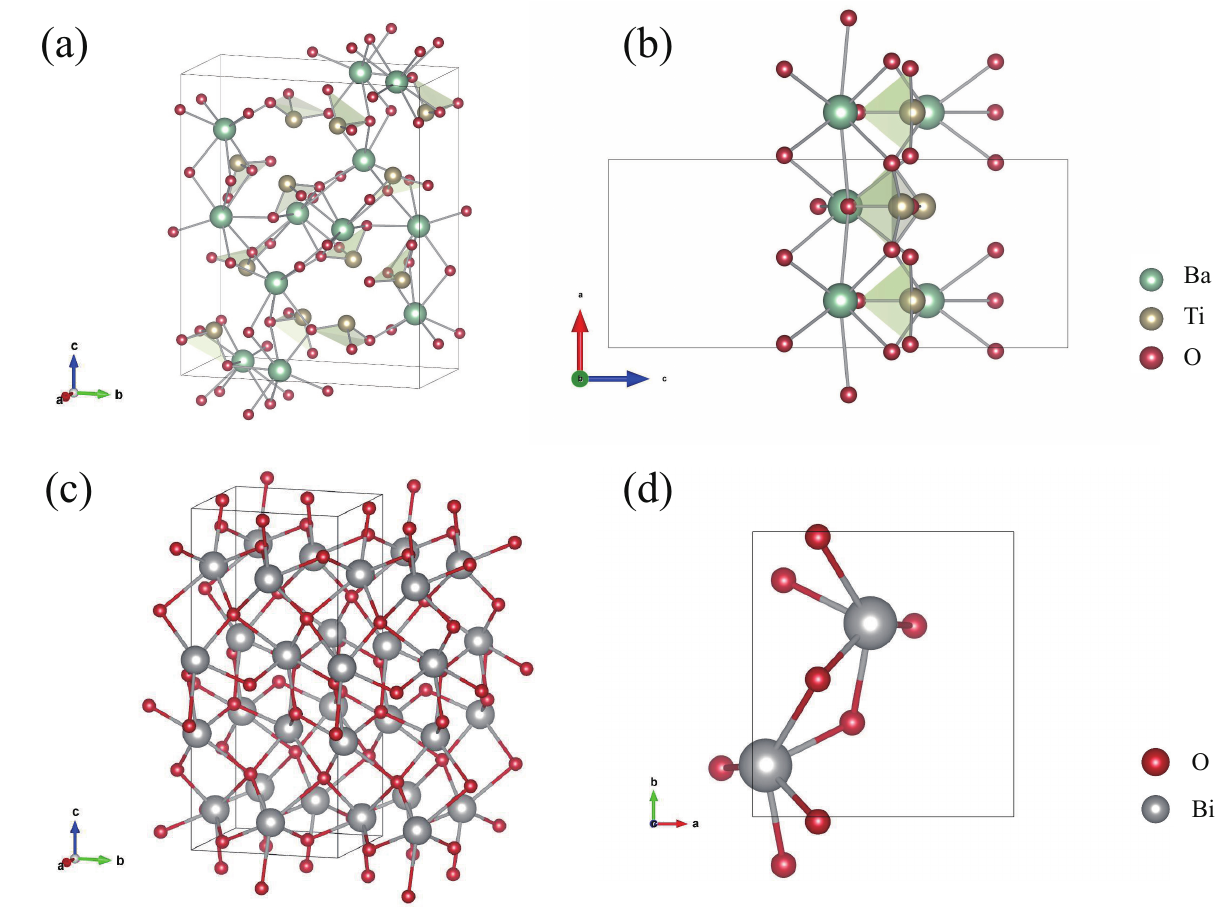}
    \caption{Crystal structures and phonon bands of (a,b) BaTeO$_3$ and (c,d) Bi$_2$O$_3$. }
    \label{fig:structures}
\end{figure}

To demonstrate the performance of the PARCE model we proposed, two crystals from MP\cite{jain2013commentary}, i.e. BaTeO$_3$(MP-556021) and Bi$_2$O$_3$ (MP-556549), were selected for detailed analysis, without loss of geneneity. The two crystals of BaTeO$_3$ with sixty atoms and Bi$_2$O$_3$ with twenty atoms in the primitive cell, respectively, are shown in Figure~\ref{fig:structures}. Crystallographically, as shown in Figure~\ref{fig:structures}(a,b), BaTeO$_3$ belongs to the Orthorhombic crystal system, with a space group of Pnma, and its lattice constants of the unit cell are $a =  6.237$~\AA, $b = 12.516$~\AA, and $c = 15.193$~\AA, with lattice angles $\alpha = 90^\circ$, $\beta = 90^\circ$, and $\gamma = 90^\circ$. As shown in Figure~\ref{fig:structures}(c,d), the crystal structure of Bi$_2$O$_3$ belongs to the orthorhombic system with space group Pccn, and its unit-cell parameters are $a = 5.06$~\AA, $b = 5.57$~\AA, and $c = 12.63$~\AA, with the angles $\alpha = \beta = \gamma = 90^\circ$.

By using the pymatgen package\cite{ong2013python}, five key structural parameters for crystal materials can be extracted and calculated. For BaTeO$_3$, the obtained parameters are as follows: a space group of 62, a Pearson symbol of oP60, a Wyckoff sequence of ``36c 24d”, a $c/a$ axial ratio of 2.42345, and a $\beta$ angle of $90^\circ$, as listed in Table~\ref{tab:parameters}.
Based on the previously described clustering scheme and the selected combination of key parameters, the reference cluster ID for crystal materials was determined using the K-means algorithm. For BaTeO$_3$, based on the combination of space group, Pearson symbol, and $c/a$ ratio, the resulting cluster ID is 29. Subsequently, by applying the Affinity Propagation clustering algorithm to the representative phonon information of each cluster, Cluster 29 was further classified into the fifth category. Consequently, the reference label used when training the RCNet-ensemble model for BaTeO$_3$ is ``Cluster 29 (Category 4).”
%
By using PHONOPY \cite{Togo2023}, the phonon frequencies at the $\Gamma$ point for BaTeO$_3$ were obtained and are listed in Table~\ref{tab:gamma-phonons}. The phonon-frequency data length was truncated to 30 to standardize the input. After inputting the phonon frequencies into the model, the predicted structural cluster ID was 29, which is in good agreement with the calculated structural cluster ID. 

\begin{table}[ht!]
\centering
\caption{Material prediction results for BaTeO$_3$ and Bi$_2$O$_3$.}
\begin{tabularx}{0.8\textwidth}{lXX}
\hline
\textbf{Property} & \textbf{BaTeO$_3$} & \textbf{Bi$_2$O$_3$} \\ \hline
Atom Count               & 60              & 20 \\
Pearson Symbol           & oP60           & oP20 \\
Wyckoff Sequence         & 36c 24d       & 4c 16e  \\
Space Group Number       & 62             & 56 \\
$c/a$ Ratio              & 2.42342      & 2.5666 \\
$\beta$ Angle            & 90$^\circ$ & 90.00$^\circ$ \\
True Cluster ID          & 29              & 0 \\
Class ID                 & 4              & 3 \\
\hline
\end{tabularx}
\label{tab:parameters}
\end{table}


For Bi$_2$O$_3$, the five key structural parameters are extracted as follows: a space group of 56, a Pearson symbol of oP20, a Wyckoff sequence of ``4c 16e,” a $c/a$ ratio of 2.5666, and a $\beta$ angle of $90.00^\circ$. The K-means algorithm, applied to the combination of space group, Pearson symbol, and $c/a$ ratio, yielded a reference cluster ID of 0. Subsequently, the Affinity Propagation clustering algorithm based on representative phonon frequencies classified Cluster0 into the fourth category. Accordingly, the reference label for Bi$_2$O$_3$ is ``Cluster 0 (Category 3).”
In the prediction stage, the phonon frequencies at the $\Gamma$ point for Bi$_2$O$_3$ were obtained and are listed in Table~\ref{tab:gamma-phonons}.
By inputting the truncated phonon frequencies of Bi$_2$O$_3$ into the integrated model, a cluster ID of 0 was obtained, which is in good agreement with the calculated one.

\subsection{PARCE for Raman Spectroscopy}
As described in ~\ref{sec:raman}, the vibrational modes that are Raman active require the corresponding basis functions are quadratic\cite{dresselhaus2007group}, which indicates that the Raman-active modes are the subgroup of the phonon modes at the $\Gamma$-point.
%
~The dynamical matrix elements for crystal materials selected from MP database were predicted by the PHONAX package\cite{fang2024phonon}, and their corresponding second-order force constants were calculated as well. Then the phonon frequencies and their irreducible representations were calculated by using the PHONOPY package\cite{Togo2023}, followed by symmetry-based extraction of Raman-active modes according to their basis functions. 
In this way, the Raman database containing 13034 materials records were built.


\begin{figure}[ht!]
    \centering
    \includegraphics[width=0.9\linewidth]{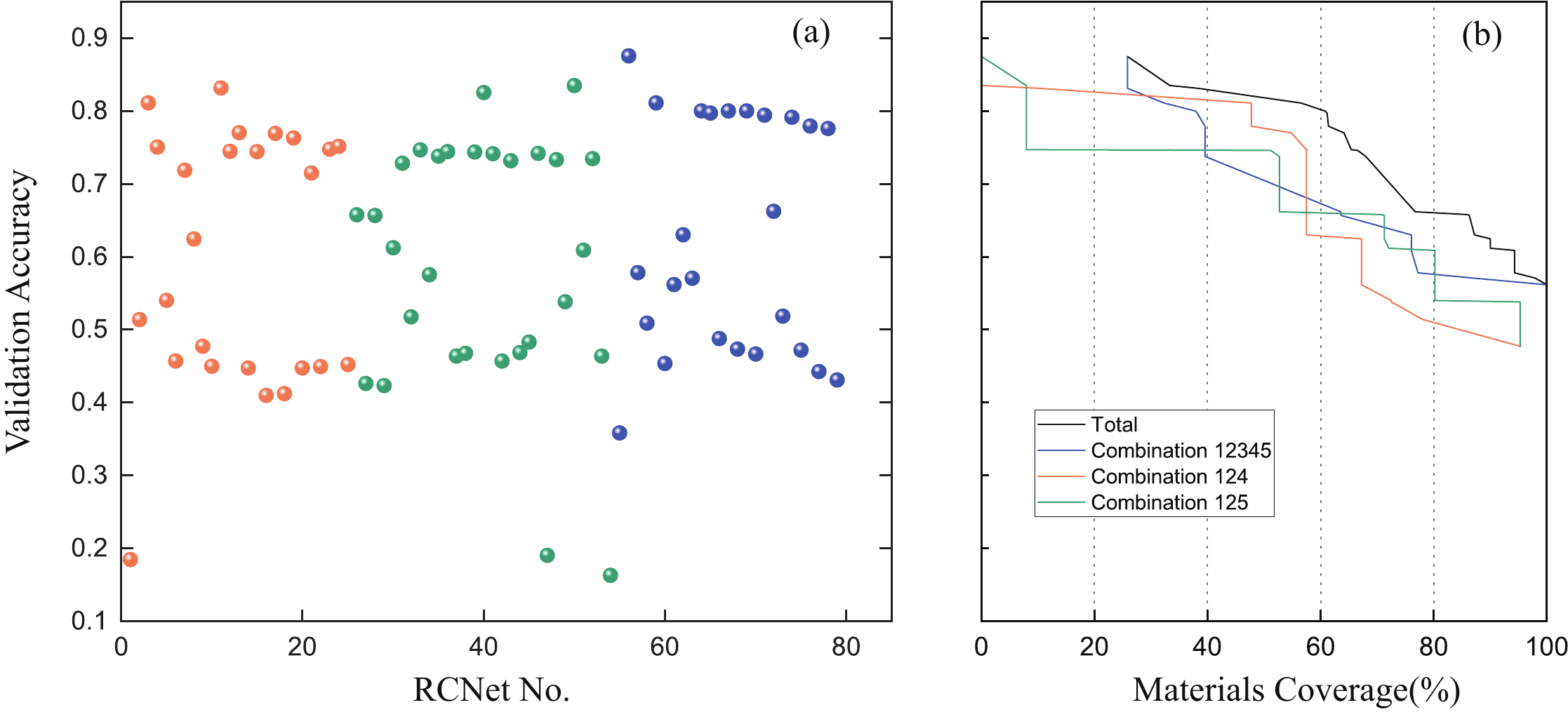}
    \caption{(a) Validation accuracy distribution across RCNet configurations based on the combinations of (124) (orange), (125) (green), and (12345) (blue). (b) Materials coverages for respective and total combinations of key pxarameters.}
    \label{fig:raman_performance}
\end{figure}

Similar to the case for phonon frequencies at the $\Gamma$ point, the Raman frequencies are truncated and standardized as well, and then the Affinity Propagation Clustering is performed to divide the dataset into $N_{lM}$ ($N_{lm} = 116$) categories, with each subset serving as the training domain for an individual RCNet model. After training, the validation accuracy corresponding to each RCNet, as well as the material coverage under different accuracy thresholds, are shown in Figure~\ref{fig:raman_performance}.
By applying a 60\% accuracy threshold to filter the RCNets results in four retained models. The training subsets used by these models cover 94.26\% of the full training dataset. On the dataset containing 13,034 materials, the ensemble model achieves an inference accuracy of 0.6265.  

\section{Conclusion}
In this study, we developed PARCE, an precision-adaptive deep learning ensemble framework. This model transcends the limitations of data dimensionality by accurately extracting structural features from phonon vibration data of varying lengths, thereby achieving robust identification of structural types in unknown compounds. Regarding the classification scheme, PARCE integrates the core crystallographic parameters—including space groups, Pearson symbols, Wyckoff sequences, and unit cell parameters (i.e. $c/a$ ratios and $\beta$ angles)—to construct hierarchical descriptors that succinctly encapsulate intrinsic crystal characteristics. The validation results demonstrate that PARCE exhibits exceptional predictive accuracy in a theoretical dataset of more than 40,000 compounds and successfully achieves reliable structural classification for more than 10,000 Raman spectra. These findings provide a high-efficiency and intelligent toolset for the automated identification of unknown phases in high-throughput experiments.


\section*{Data Availability}
The datasets generated during and/or analysed during the current study are available from the corresponding author on reasonable request.

\section*{Acknowledgements}
This work is supported by the National Key R\&D Program of China (2023YFA1608501), the Natural Science Foundation of China (12574243,92580128), and Shanghai Municipal Natural Science Foundation under Grant No. 24ZR1406600.

\section*{Competing Interests statement}

The authors declare no competing interests.

\section*{Author Contributions}
H. Z. conceived the project and contributed to securing funding. H. Z. supervised the research. HY. C. and M. D. developed and trained the neural networks and analyzed the results.  HJ. C. performed the high-throughput calculations. HY. C. and M. D. wrote the original manuscript. RL. L. and X. T. contributed to the figures.  RX. L., Y. Z., X. C. and W. L. contributed to the discussion of results and manuscript preparation.



\end{document}